\documentclass[12pt,dvips]{article}
\usepackage[dvips]{graphicx}
\usepackage{graphicx}

\graphicspath{{plots/}}
\DeclareGraphicsExtensions{.eps.gz,.eps,.ps,.ps.gz}

\parskip=\itemsep               
\newcommand{\lwig}{\mbox{\,\raisebox{.3ex}
    {$<$}$\!\!\!\!\!$\raisebox{-.9ex}{$\sim$}\,}}
\newcommand{\gwig}{\mbox{\,\raisebox{.3ex}
    {$>$}$\!\!\!\!\!$\raisebox{-.9ex}{$\sim$}}\,}
\newcommand{\lambdabar}{{\hbox{$\lambda_e$\kern-1.9ex\raise+0.45ex\hbox{--}
\kern+0.2ex}}}

\newif\ifhepph

\hepphtrue

\ifhepph\date{\empty}\fi

\title{
\ifhepph{\normalsize\rightline{WUB 03-03}\rightline{ITP-BUDAPEST 593}\rightline{DESY 03-022}\rightline{hep-ph/0303080}}\fi
\vskip 1cm 
\bf\boldmath
Electroweak instantons as a solution to the ultrahigh energy cosmic ray puzzle
       \vspace{21mm}} 
\author{
Z.~Fodor$^{a,b}$, S.~D.~Katz$^c$\thanks{On leave from Institute for Theoretical Physics, E\"otv\"os University, 
Budapest, Hungary.}, A.~Ringwald$^c$, and H.~Tu$^c$\\[1cm]
\it $^a$Department of Physics, University of Wuppertal,  
Germany\\
\it $^b$Institute for Theoretical Physics, E\"otv\"os University, 
Budapest, Hungary\\
\it $^c$Deutsches Elektronen-Synchrotron DESY, 
Hamburg, Germany
}

\begin{document}
\begin{titlepage} 
  \maketitle
\begin{abstract}
We propose a scenario in which a simple power-like primary spectrum
for protons with sources at cosmological distances 
leads to a quantitative description of all
the details of the observed cosmic ray spectrum for energies 
from $10^{17}$~eV to $10^{21}$~eV. As usual, the ultrahigh energy protons
with energies above $E_{\rm GZK}\approx 4\cdot 10^{19}$~eV loose a large fraction of their energies
by the photoproduction of pions on the cosmic microwave background, 
which finally decay mainly into neutrinos. In our scenario, 
these so-called cosmogenic neutrinos interact with 
nucleons in 
the atmosphere 
through Standard Model electroweak instanton-induced processes and produce air showers which are 
hardly distinguishable from ordinary hadron-initiated air 
showers. In this way, they give rise to 
a second contribution to the observed cosmic ray spectrum -- in addition to the one 
from above mentioned protons -- which reaches beyond $E_{\rm GZK}$.    
Since the whole observed spectrum is uniquely determined by a single primary injection spectrum, 
no fine tuning is needed to fix the ratio of the spectra below and above $E_{\rm GZK}$.
The statistical analysis shows an excellent goodness of this scenario. 
Possible tests of 
it range from observations at cosmic ray facilities and neutrino
telescopes to searches for QCD instanton-induced processes at HERA. 
\end{abstract}


\thispagestyle{empty}
\end{titlepage}
\newpage \setcounter{page}{2}

\section{\label{intro}Introduction}

The puzzle of ultrahigh energy cosmic rays (UHECRs) 
is about $40$ years old. About twenty mysterious events were 
observed above 10$^{20}$~eV
by five different air shower observatories  
(AGA\-SA~\cite{Takeda:1998ps}, Fly's Eye~\cite{Bird:yi
}, Haverah Park~\cite{Lawrence:cc
}, HiRes~\cite{Abu-Zayyad:2002ta
}, and Yakutsk~\cite{Efimov91}; 
for reviews, see Ref.~\cite{Nagano:ve
}).
Though some small-angle clustering in the arrival direction of the 
UHECRs is observed, the overall event distribution is
isotropic.
This indicates that they originate from several sources.  
No source is known, however, within a distance of $50$~Mpc. This is 
rather peculiar, since $50$~Mpc is the 
characteristic distance ultrahigh energy nucleons travel
before they loose a large fraction of their energy. 
A sharp drop around 
the Greisen-Zatsepin-Kuzmin (GZK) cutoff
$E_{\rm GZK}$$\approx 4\cdot 10^{19}$~eV is therefore predicted in the
cosmic ray spectrum~\cite{Greisen:1966jv
}.
The available data show no such drop\footnote{There is an ongoing
debate whether the excess of events above $E_{\rm GZK}$ is 
significant and whether the data from different collaborations are
mutually consistent~\cite{Bahcall:2002wi
}. We will comment on this point below.}. 

The reason for the expected drop is a well established 
elementary process.
Above $E_{\rm GZK}$, protons produce pions through the 
interaction with photons from the 
$2.7$~K cosmic microwave background (CMB). The produced pions
decay, resulting in the so-called cosmogenic 
neutrinos~\cite{Beresinsky:qj}.
The attenuation length of protons above the 
GZK cutoff is about $50$~Mpc.
The basic question is: 
if the sources of ultrahigh energy cosmic rays are indeed at cosmological distances, 
how could they reach us with energies above $10^{20}$~eV? 
No conventional explanation is known for this question. 

At the relevant energies among the known particles only neutrinos can propagate without 
significant energy loss from cosmological distances to us. 
It is this fact which led, on the one hand, 
to scenarios invoking 
hypothetical -- beyond the Standard Model --  
strong interactions of ultrahigh energy 
cosmic 
neutrinos~\cite{Beresinsky:qj} and, on the other hand, to the Z-burst 
scenario~\cite{Fargion:1997ft,
Fodor:2001qy,
Kalashev:2001sh}.

In the latter, ultrahigh energy cosmic neutrinos (UHEC$\nu$s) 
produce Z-bosons through annihilation with the relic neutrino background from the big bang.
On earth, we observe the air showers initiated by the protons and photons 
from the hadronic decays of these Z-bosons. Though the required ultrahigh energy
cosmic neutrino flux 
is smaller than present upper bounds, it is not easy to 
conceive a production mechanism yielding 
a sufficiently large one. In the near future, the neutrino telescopes AMANDA~\cite{Andres:2001ty}
and RICE~\cite{Kravchenko:2001id}, as 
well as the Pierre Auger Observatory~\cite{Zavrtanik:2000zi} for extensive air showers, 
can directly confirm or exclude this scenario.

Scenarios based on strongly interacting neutrinos use the fact
that the observed cosmic ray flux above $E_{\rm GZK}$ can be fairly well described
by the 
predicted~\cite{Yoshida:pt,
Kalashev:2002kx} cosmogenic 
neutrino flux. 
In these scenarios,  neutrinos with energies 
above $\approx 10^{20}$~eV originating from the GZK process are assumed to 
have a large cross-section for the scattering off nucleons 
and to initiate extensive
air showers high up in the atmosphere, like hadrons.   
This is usually ensured by new types of TeV-scale interactions beyond the Standard Model,
such as 
arising through gluonic bound state leptons~\cite{Bordes:1997bt
},
TeV-scale grand unification with leptoquarks~\cite{Domokos:2000dp}, 
or Kaluza-Klein modes from extra 
compactified 
dimensions~\cite{Domokos:1998ry
} 
(see, however, Ref.~\cite{Kachelriess:2000cb
}); 
for earlier and further proposals, see Refs.~\cite{Domokos:1986qy} and 
\cite{Barshay:2001eq
}, respectively. 
Until now, none of these ideas have direct 
experimental verification.

In this Letter, 
we propose another strongly interacting
neutrino scenario to solve the GZK problem,  
which -- in contrast to previous proposals -- 
is based entirely on the Standard Model of particle physics. 
It exploits non-perturbative electroweak instanton-induced 
processes
for the interaction of cosmogenic neutrinos with nucleons in the atmosphere, 
which may have a sizeable cross-section above a threshold energy 
$E_{\rm th}={\mathcal O}( (4\pi M_W/\alpha_W )^2)/(2 m_p) = {\mathcal O}( 10^{18})$~eV, where 
$M_W$ denotes the W-boson mass and $\alpha_W$ the electroweak fine structure 
constant~\cite{Aoyama:1986ej,
Morris:1993wg,Ringwald:2002sw}.  
For the first time in the literature, we present  a 
detailed statistical analysis of the agreement between observations and 
predictions from strongly interacting 
neutrino scenarios.   
 
Our scenario can be summarized as follows. We assume  a 
standard power-like 
primary spectrum for protons injected from sources at cosmological distances, 
which extends beyond the
GZK cutoff. 
After propagation through the CMB, the protons -- arriving at earth 
mostly with energies below 
$E_{\rm GZK}$ -- will be one component of the 
observed cosmic ray spectrum.  
The spectrum of the produced cosmogenic neutrinos 
is entirely determined by the proton injection spectrum and can therefore be
determined precisely, including all known effects. 
The cosmogenic neutrinos travel unaffected through the CMB. However, for energies above
$\approx 10^{19}$~eV, they have a large 
cross-section for interactions with nuclei in the atmosphere due to electroweak instanton-induced processes. 
They give rise, therefore, to a second, predictable component of the observed cosmic ray spectrum, which dominates 
above the GZK cutoff over the first, proton-initiated component.
Our proposal leads to an explanation of the observed cosmic ray spectrum
simultaneously above and below the GZK 
cutoff, without the need to fix the ratio of the fluxes below and above 
$E_{\rm GZK}$  by hand\footnote{This feature is shared with all strongly interacting neutrino scenarios. 
In contrast to these other scenarios,  
in our case, however, the threshold energy is automatically fixed by Standard Model parameters ($M_W$ and 
$\alpha_W$).}, as it is necessary in most alternative 
proposals.
The goodness of the scenario
is studied by statistical methods and an excellent agreement
is seen between the predictions and the observations.

Our analysis proceeds in three steps, which are
performed in Sects.~\ref{fluxes}, \ref{inst-spect}, and \ref{comparison}.
{\em i)} 
First, we study the consequences of a power-like proton 
injection spectrum. 
We determine the resulting proton and neutrino fluxes on earth, 
taking into account the appropriate types of energy losses.  
{\em ii)} In the second step, we 
calculate the spectrum of cosmogenic neutrino-initiated electroweak 
instanton-induced air showers.  
{\em iii)} The third step consists in the 
comparison of the observed UHECR spectrum with the prediction arising from
an inclusion of instanton-induced processes. 
Based on the goodness of the scenario, we determine 
the confidence region in the parameter space of our
scenario. Finally, we summarize our result and present our conclusions in Sect.~\ref{conclusions}. 

\section{\label{fluxes}Proton and cosmogenic neutrino fluxes}

We start with a power-like 
injection spectrum per co-moving volume  of protons with energy $E_i$, spectral index $\alpha$, 
and redshift ($z$) evolution index $m$,
\begin{equation}
j_p =j_0\,E^{-\alpha}_i\,\left(1+z\right)^m\,\theta(E_{\rm max}-E_i)\,
\theta(z-z_{\rm min})\,\theta(z_{\rm max}-z)\,.
\end{equation}
Here, $j_0$ is a normalization factor, $E_{\rm max}$ is
the maximal energy, which can be reached through 
astrophysical 
accelerating processes in a bottom-up 
scenario, and $z_{\rm min/max}$ take into account that nearby/very early there are
no astrophysical sources.     
As we will see in our comparison with UHECR data in Sect.~\ref{comparison}, 
the overall normalization $j_0$ is fixed by
the observed flux, 
and our predictions are quite insensitive to the
specific choice for 
$E_{\rm max}$, $z_{\rm min}$, and $z_{\rm max}$, within their anticipated
values. 
The main sensitivity arises from the spectral parameters $\alpha$ and $m$, 
for which we determine the 1- and 2-sigma confidence regions in Sect.~\ref{comparison}.

The injected protons propagate through the CMB. 
This propagation can be 
described~\cite{Bahcall:1999ap
}
by $P_{b|a} (r,E_i;E )$ functions, which give the expected number of 
particles of type $b$ 
above the threshold energy $E$ if one particle of type $a$ 
started at  a distance $r$ with energy $E_i$. 
With the help of these propagation functions, 
the differential flux of protons ($b=p$) 
and cosmogenic neutrinos 
($b=\nu_i, \bar\nu_i$) at earth, 
i.e. their number $N_b$ arriving at earth with energy $E$ per units of energy, 
area ($A$), time ($t$) and solid angle ($\Omega$), 
can be expressed as   
\begin{equation}
\label{flux-earth}
F_{b} ( E ) \equiv \frac{{\rm d}^4 N_{b}}{{\rm d}E\,{\rm d}A\,{\rm d}t\,{\rm d}\Omega}
=
 \int_0^\infty {\rm d}E_i \int_0^\infty {\rm d}r 
\,\left(1+z(r)\right)^{3}
\,(-)\frac{\partial P_{b|p}(r,E_i;E)}{\partial E}
\,j_p (r,E_i)\,.
\end{equation}
In our analysis we go, according to ${\rm d}z = - (1+z)\,H(z)\,{\rm d}r/c$,  
out to distances $R_{\rm max}$ 
corresponding to $z_{\rm max} = 2$ (cf.~Ref.~\cite{Waxman:1995dg}), while 
we choose $z_{\rm min}=0.012$ in order to take into account the fact that within 
$50$~Mpc there are no astrophysical sources of UHECRs. 
We use the expression 
\begin{equation}
\label{H-Omega}
H^2(z) = H_0^2\,\left[ \Omega_{M}\,(1+z)^3 
+ \Omega_{\Lambda}\right]
\end{equation} 
for the relation of the Hubble expansion rate at redshift $z$ to the present one.
Uncertainties of the latter, $H_0=h$ 100 km/s/Mpc, with 
$h=(0.71\pm 0.07)\times^{1.15}_{0.95}$~\cite{Hagiwara:fs}, 
are included. 
In Eq.~(\ref{H-Omega}), $\Omega_{M}$ and $\Omega_{\Lambda}$, with $\Omega_M+\Omega_\Lambda =1$, are the present 
matter and vacuum energy densities in terms of the critical density. As default values we choose
$\Omega_M = 0.3$ and $\Omega_\Lambda = 0.7$, as favored today. Our results
turn out to be pretty insensitive to the precise values of the cosmological parameters.

We calculated $P_{b|a}(r,E_i;E)$ in two steps. 
{\em i)} First, the SOPHIA
Monte-Carlo program~\cite{Mucke:1999yb} was 
used for the simulation of photohadronic processes of protons with the CMB photons. 
For $e^+e^-$ pair production we used the continuous energy loss approximation, since the
inelasticity is very small ($\approx 10^{-3}$).
We calculated
the $P_{b|a}$ functions for ``infinitesimal'' steps ($1\div 10$~kpc) as a function of 
the redshift $z$.
{\em ii)} 
We multiplied the corresponding infinitesimal probabilities  
starting at a distance $r(z)$ down to earth with $z=0$. 
The details of the calculation of the $P_{b|a}(r,E_i;E )$ functions for
protons, neutrinos, charged leptons, and photons will be published
elsewhere~\cite{P:xxx}.

The determination of the propagation
functions took approximately 
one day on an average personal computer.
In this connection, the advantage of the formulation of the 
spectra~(\ref{flux-earth}) in terms of the propagation functions becomes
evident. The latter have to be determined only once and for all. 
Without the use of the propagation functions, one would have to perform
a simulation for any variation of the input spectrum 
($\alpha,m,\ldots $),  
which requires excessive computer power.
Since the propagation functions are of universal usage, we decided
to make the latest versions of $-\partial P_{b|a}/\partial E$ 
available for the public via the World-Wide-Web URL 
www.desy.de/\~{}uhecr \,.

As a check on our propagation functions, we have compared our predictions for the 
spectra~(\ref{flux-earth}) with the ones presented in Ref.~\cite{Kalashev:2002kx} 
for some specific values of the spectral parameters 
($\alpha$, $m$, $E_{\rm max}$, $z_{\rm min}$, $z_{\rm max}$, $j_0$) and found quite good agreement.

\section{\label{inst-spect}Spectrum of instanton-induced air showers}

In this section, we 
exploit a recent prediction of the electroweak instanton-induced parton-parton 
cross-section~\cite{Ringwald:2002sw} and  
determine the spectrum of instanton-induced air showers, which
are initiated by the cosmogenic neutrino flux~(\ref{flux-earth}) impinging on the earth's 
atmosphere. 

Let us start with a review of the current knowledge about electroweak instantons.
In the Standard Model of electroweak interactions (Quantum Flavor Dynamics (QFD))
there are certain processes  which fundamentally can not be described by ordinary
perturbation theory. These processes 
are associated with axial 
anomalies and 
manifest themselves as anomalous violation of baryon plus lepton  number 
($B+L$)~\cite{'tHooft:up
}. They are induced by topological fluctuations of the non-Abelian gauge
fields, notably by instantons~\cite{Belavin:fg
}.
In Minkowski space-time, instantons  
describe tunneling transitions between degenerate, topologically
inequivalent vacua.
The corresponding tunneling barrier
is given by the energy of
the sphaleron~\cite{Klinkhamer:1984di}, an 
unstable static solution of the
Yang-Mills equations,  and of order 
$M_{\rm sp} \approx \pi M_W/\alpha_W \approx 10$~TeV. The corresponding processes 
violate $B+L$ according to the selection rule 
$\triangle  B = \triangle L =-3$. 

\begin{figure}
\vspace{-1.6cm}
\begin{center}
\includegraphics*[width=8.3cm,clip=]{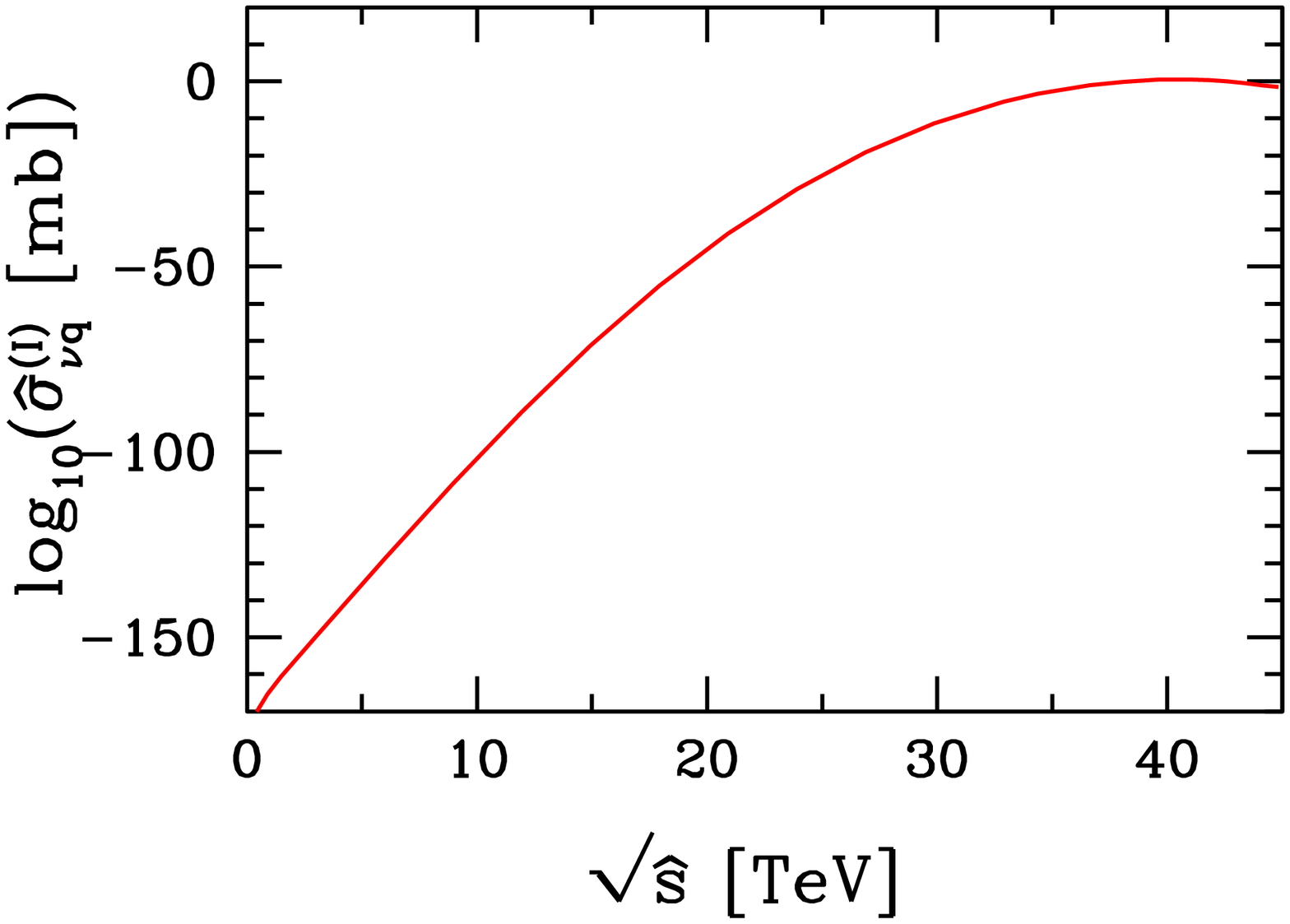}
\includegraphics*[width=8.3cm,clip=]{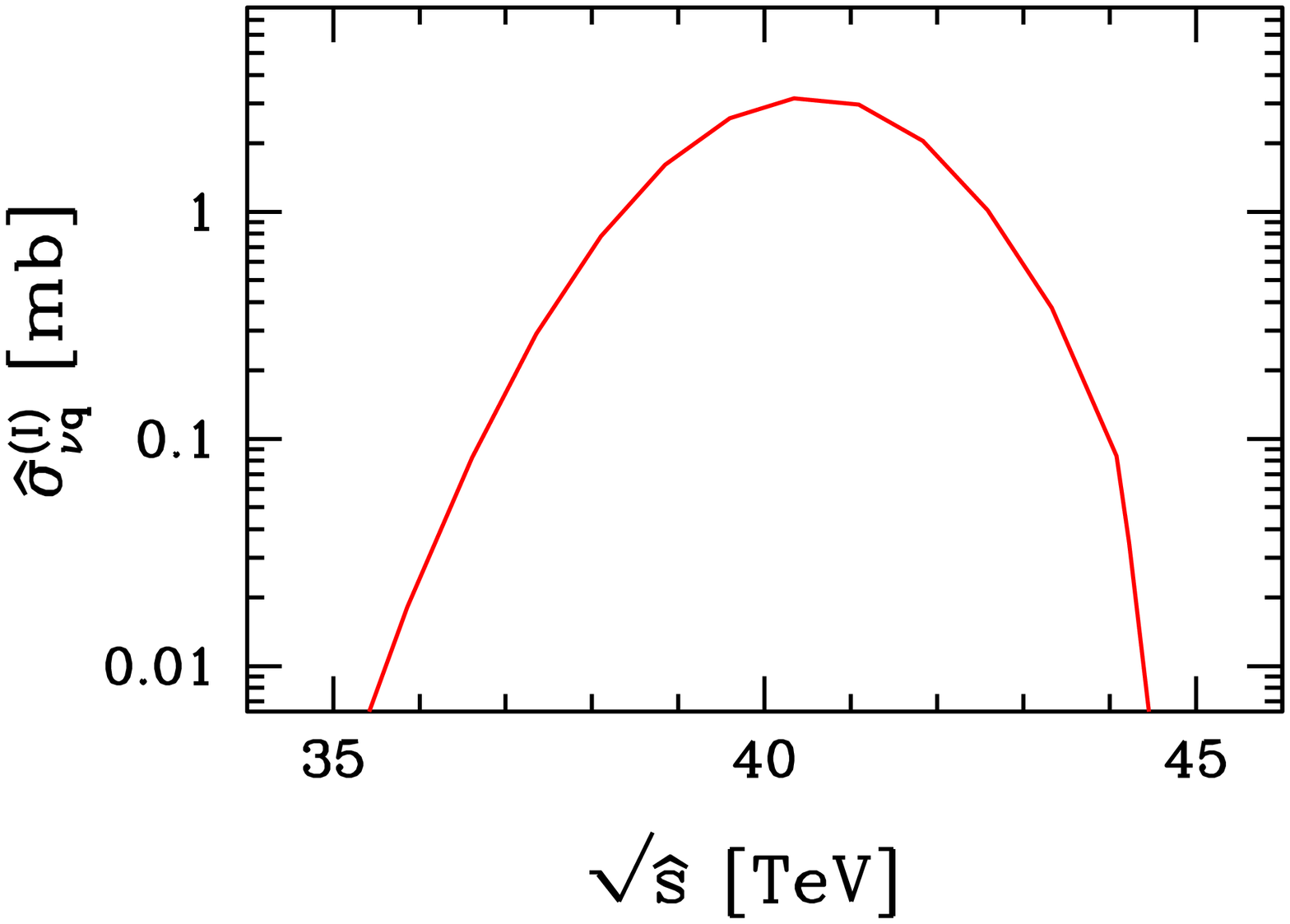}
\vspace{-1.0cm}
\caption[dum]{The electroweak instanton-induced neutrino-quark cross-section $\hat\sigma_{\nu q}^{(I)}$
as predicted in Ref.~\cite{Ringwald:2002sw} for the whole range of the CM energy $\sqrt{\hat s}$ 
(left) and near the maximum (right).  
\label{cross-parton}}
\end{center}
\end{figure}

It is generally accepted that such topological fluctuations and
the associated $B+L$ violating processes are very important at high 
temperatures~\cite{Kuzmin:1985mm
} and have therefore a crucial impact on the evolution of the baryon and lepton
asymmetries of the 
universe\footnote{Standard Model 
electroweak baryogenesis seems excluded, however, due to the weakness of the 
electroweak phase 
transition~\cite{Arnold:1992rz
} (for reviews, see Refs.~\cite{Rubakov:1996vz,Laine:2000xu
}), while thermal leptogenesis~\cite{Fukugita:1986hr,Buchmuller:2002xm} is quite successful.}. 
It is, however, still debated whether manifestations of such 
fluctuations -- involving notably the associated production of 
${\mathcal O}(1/\alpha_W) \approx 30$ 
W/Z-bosons in addition to the anomalously produced quarks and leptons -- 
may be directly observed in high-energy scattering 
processes~\cite{Aoyama:1986ej
}.
Despite considerable 
theoretical~\cite{McLerran:1989ab
} and 
phenomenological~\cite{Morris:1993wg,Farrar:1990vb
} efforts, the actual
size of the cross-sections in the relevant, tens of TeV energy regime was never unanimously established (for  
reviews, see Refs.~\cite{Rubakov:1996vz,Mattis:1991bj
}).

There is a close analogy~\cite{Balitsky:1993jd
} between QFD and hard QCD instanton-induced processes 
in deep-inelastic 
scattering~\cite{Moch:1996bs
}.
Recent information about the latter 
-- both from lattice simulations~\cite{Smith:1998wt
} and from the H1 experiment at HERA~\cite{Adloff:2002ph} -- has been used by one of the authors to 
learn about the fate of electroweak $B+L$ violation and associated multi-W/Z production 
at high energies~\cite{Ringwald:2002sw} (for a review, see Ref.~\cite{Ringwald:2003px}). 
The prediction for the electroweak instanton-induced neutrino-quark cross-section 
$\hat\sigma_{\nu q}^{(I)}$ is 
displayed in Fig.~\ref{cross-parton} as a function of the neutrino-quark center-of-mass (CM) energy
$\sqrt{\hat s}$. At small CM energies, 
the cross-section is really tiny, e.g. $\hat\sigma_{\nu q}^{(I)}\approx 10^{-141}$~pb at 
$\sqrt{\hat s}\approx 3$~TeV, but steeply growing. 
Nevertheless, it stays unobservably small, 
$\hat\sigma_{\nu q}^{(I)}\lwig 10^{-26}$~pb for $\sqrt{\hat s}\,\lwig\, 22.5$~TeV, 
in the quite conservative fiducial
kinematical region inferred via the QFD--QCD analogy 
from lattice data and HERA. It was noted that a slight extrapolation towards
larger energies -- still compatible with lattice results and HERA -- 
points to a cross-section $\approx 10^{-6}$~pb at
a CM energy of about $30$~TeV, which is within the reach of  the Very Large Hadron Collider.

\begin{figure}
\vspace{-1.6cm}
\begin{center}
\includegraphics*[width=8.3cm,clip=]{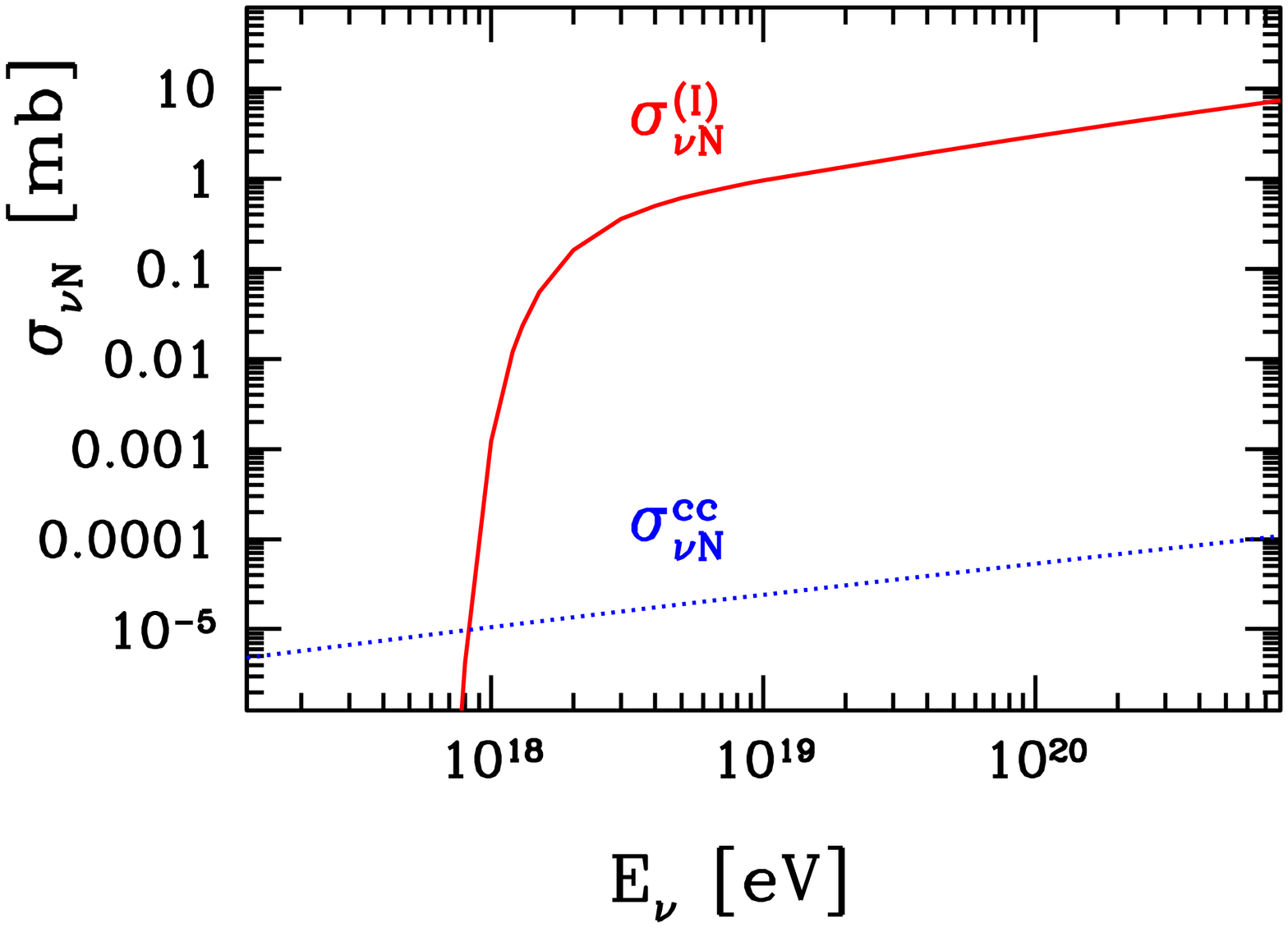}
\includegraphics*[width=8.3cm,clip=]{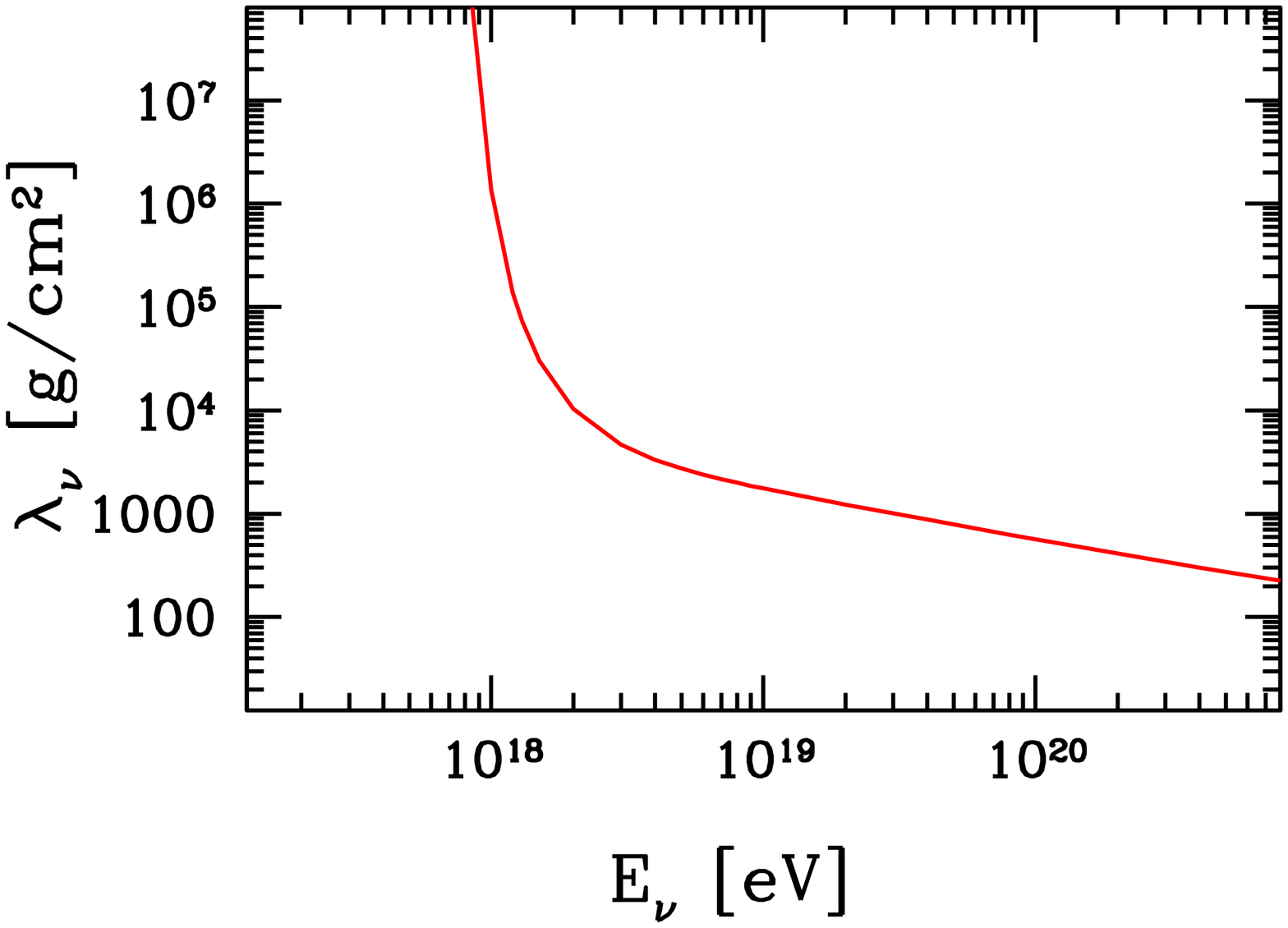}
\vspace{-1.0cm}
\caption[dum]{
{\em Left:} 
Prediction of the electroweak instanton-induced neutrino-nucleon cross-section 
$\sigma_{\nu N}^{(I)}$ (solid) in comparison with the charged current cross-section 
$\sigma_{\nu N}^{\rm cc}$ (dotted) from Ref.~\cite{Gandhi:1998ri}, as a function of 
the neutrino energy $E_\nu$ in the nucleon's rest frame. 
{\em Right:} Neutrino interaction length due to combined effects of charged current
interactions and instanton-induced processes.
\label{cross-nuN}}
\end{center}
\end{figure}

In this Letter, we will 
use the prediction from Ref.~\cite{Ringwald:2002sw} even at higher 
energies\footnote{It should be kept in mind, however, that, at the energies
of interest here, the prediction in Fig.~\ref{cross-parton} is rather an 
educated extrapolation or guess (cf. Ref.~\cite{Ringwald:2002sw}).}, up to and above 
$\sqrt{\hat s}\approx 40$~TeV, where the cross-section reaches its maximum
of order a few  
millibarn (cf. Fig.~\ref{cross-parton} (right)). 
The corresponding neutrino-nucleon cross-section is obtained after folding the 
parton cross-section $\sigma_{\nu q}^{(I)}$ with the quark density functions $f_q$, 
\begin{equation}
\label{sig_nuN_i_pdf}
\sigma_{\nu N}^{(I)} (s) =
\sum_{q} \int^1_0
{\rm d}x\,f_q (x,\mu )\,\hat \sigma_{\nu q}^{(I)} (xs)
\, ,
\end{equation}
where $s$ denotes the neutrino-nucleon CM energy squared and $\mu$ the factorization scale. 
For our numerical integration we have used various sets of parton distributions as they are 
implemented in the parton distribution library 
PDFLIB~\cite{Plothow-Besch:1993qj
}. Uncertainties associated with different parton distribution sets 
are in the ${\mathcal O}(20)\, \%$ range and are not explicitely shown in the following.
Figure~\ref{cross-nuN} (left) displays the prediction of the electroweak instanton-induced
neutrino-nucleon cross-section as a function of the neutrino energy $E_\nu$ in the nucleon's rest frame 
for a choice $\mu =M_W$ of the factorization scale.  
Above a threshold at about $E_\nu\approx 10^{18}$~eV, it quickly reaches 
one millibarn at about $10^{19}$~eV, and tends to 
grow power-like, due to the growth of the sea quark distributions in the nucleon at small $x$,  
quite analogous to the standard charged current cross-section 
$\sigma_{\nu N}^{\rm cc}$ (cf. Fig.~\ref{cross-nuN} (left)).
It should be noted that such a cross-section will lead, via dispersion relations, to 
lower energy deviations of Standard Model predictions for elastic scattering from their perturbative 
values~\cite{Rubakov:1996vz}. 
However, it is easily checked that, for the one shown in Fig.~\ref{cross-nuN} (left), 
these corrections will be unobservably small in the energy regime available at present 
accelerators~\cite{Goldberg:1998pv}.

The corresponding neutrino interaction length 
$\lambda_\nu \equiv m_p/\sigma_{\nu N}^{\rm tot}$, with 
$\sigma_{\nu N}^{\rm tot}=\sigma_{\nu N}^{\rm cc}+ \sigma_{\nu N}^{(I)}$, is shown in 
Fig.~\ref{cross-nuN} (right). 
It falls below $X_0=1031$~g/cm$^2$ -- the vertical depth 
of the atmosphere at sea 
level\footnote{\label{atm_depth}For our numerical calculations involving the atmospheric depth $X(\theta )$ 
we have used a parametrization of the US Standard Atmosphere (1976) from 
Ref.~\cite{Gandhi:1995tf}.}  -- for $E_\nu\gwig 3\cdot 10^{19}$~eV. 
The apparent 
success of our scenario is based on the unexpected
coincidence of this scale and $E_{\rm GZK}$.
Above this energy, the atmosphere becomes opaque to cosmogenic neutrinos and all of them will end up as 
air showers. Quantitatively, this fact can be described by   
\begin{equation}
\label{shower-rate}
F^{(I)} (E) \equiv 
\frac{{\rm d}^4 N^{(I)}}{{\rm d}E\,{\rm d}t\,{\rm d} A\,{\rm d}\Omega}
=
\frac{\sigma_{\nu N}^{(I)}(E)}{\sigma_{\nu N}^{\rm tot}(E)}\,F_\nu (E)\,
\left[ 1 - {\rm e}^{- \frac{X(\theta )}{\lambda_\nu (E)}}\right] 
\,,
\end{equation}
which gives 
the spectrum 
of neutrino-initiated instanton-induced air showers, 
for an incident cosmogenic neutrino flux  
$F_\nu =\sum_i [F_{\nu_i}+F_{\bar\nu_i}]$ 
from Eq.~(\ref{flux-earth}), 
in terms of the atmospheric depth$^{\ref{atm_depth}}$ $X(\theta )$,  
with $\theta$ being the zenith angle. 
For $E_\nu\gwig 3\cdot 10^{19}$~eV, one has $\lambda_\nu (E_\nu )< X_0$,
and the spectrum~(\ref{shower-rate}) 
quickly equals the incident cosmogenic neutrino flux, $F_\nu (E)$.
For $E_\nu\lwig 4\cdot 10^{18}$~eV, on the other hand, the cross-section 
$\sigma^{\rm tot}_{\nu N}(E_\nu ) \lwig 0.56$~mb corresponds to a neutrino interaction length 
$\lambda_\nu (E_\nu )\gwig 3000$~g/cm$^2$, which is comparable 
to the atmospheric depth at larger zenith angles, $\theta \gwig 70^{\circ}$. 
Therefore, for these energies, neutrino-initiated electroweak instanton-induced
showers can be searched for at cosmic ray facilities by looking for 
quasi-horizontal air showers, $\theta \gwig 70^{\circ}$~\cite{Morris:1993wg}. 
At the end of Sect.~\ref{comparison}, we will show that the rate from our prediction~(\ref{shower-rate}) is 
consistent with observational constraints found by the Fly's Eye~\cite{Baltrusaitis:mt}
and AGASA~\cite{Yoshida:2001icrc} collaborations. 

Equation~(\ref{shower-rate}) does not account
for the efficiency of an air shower array to trigger on low altitude air showers. 
Below $10^{19}$~eV, neutrino-induced showers may be initiated so close to the array that the 
showers do not spread out sufficiently to trigger the array. 
As discussed in Ref.~\cite{Morris:1993wg}, 
one may suppose that an array does not trigger on showers initiated within $X_{\rm tr}=500$~g/cm$^2$ 
of the detection level. This can
be implemented in Eq.~(\ref{shower-rate}) by replacing $X(\theta )$ with $(X(\theta )-X_{\rm tr})$.
Such an assumption seems reasonable for vertical showers seen by a ground array (AGASA), 
but is somewhat pessimistic for showers at larger zenith angles or for fluorescence detectors (HiRes).
We have performed our fit in Sect.~\ref{comparison} with/without such a ``trigger'' cut for AGASA/HiRes data. 
Its effect, however, turned out to be negligible. 

Proton-initiated electroweak instanton-induced air showers have been quite 
intensively studied in Ref.~\cite{Morris:1993wg} and compared to generic proton- or iron-initiated 
air showers\footnote{On account of $\sigma_{pN}^{(I)}\ll \sigma_{pN}^{\rm gen}\approx 100$~mb, 
where $\sigma_{pN}^{\rm gen}$ is the cross-section for generic
proton-nucleon processes, the contribution of proton-initiated 
instanton-induced air showers to the UHECR spectrum can be safely ignored.}. 
While identifiable systematic differences between average showers of different 
origin could be found, the differences did not appear to be sufficient to discriminate between 
proton-initiated instanton-induced showers and fluctuations in generic showers. 
The same is expected for neutrino-initiated instanton-induced air showers, as long
as the first interaction occurs sufficiently high in the atmosphere, at a depth 
$\lwig 500$~g/cm$^2$, which happens in our case for $E_\nu\gwig 10^{20}$~eV. 
We will find from our fits in the next section that the contribution of cosmogenic neutrino-initiated
air showers to the UHECR spectrum starts to dominate at around this energy over the
proton-initiated generic component. 

\section{\label{comparison}Comparison with UHECR data}

In this section, we compare the predicted air shower spectrum from Eqs.~(\ref{flux-earth}) and 
(\ref{shower-rate}), the latter averaged over the appropriate range of zenith angles $\theta$, 
\begin{equation}
\label{flux-pred}
F_{\rm pred} (E; \alpha , m, E_{\rm max},  z_{\rm min}, z_{\rm max}, j_0 ) 
= F_{p} ( E; \ldots  ) + F^{(I)} (E; \ldots )\,, 
\end{equation}
with the observations. We perform a detailed statistical analysis
and present a measure for the goodness of the instanton-induced scenario. 

The analysis consists of two parts.
{\em i)} The UHECR collaborations give their 
results for the incoming flux in a binned form. Note, however,  
that the number of events in a given bin is integer and  
follows the Poisson distribution. In order to be able to give the
goodness of the instanton-induced scenario by 
statistical methods, we determine the number of experimentally 
observed events in a given energy bin by converting the published
values of the cosmic ray flux. We analyse the results from 
different experimental
settings separately and perform the UHECR analysis for the two
most recent results from the HiRes and AGASA collaborations, respectively.
{\em ii)} We determine the 1-sigma and 2-sigma confidence regions
for the parameters ($\alpha , m$) characterizing the proton injection 
spectrum (cf. Sect.~\ref{fluxes}). The method is similar to the
frequentist's analysis~\cite{Hagiwara:fs} and uses a Monte-Carlo
integration in the multi-dimensional space of bins.

\begin{figure}
\vspace{-1.6cm}
\begin{center}
\includegraphics*[width=8.3cm,clip=]{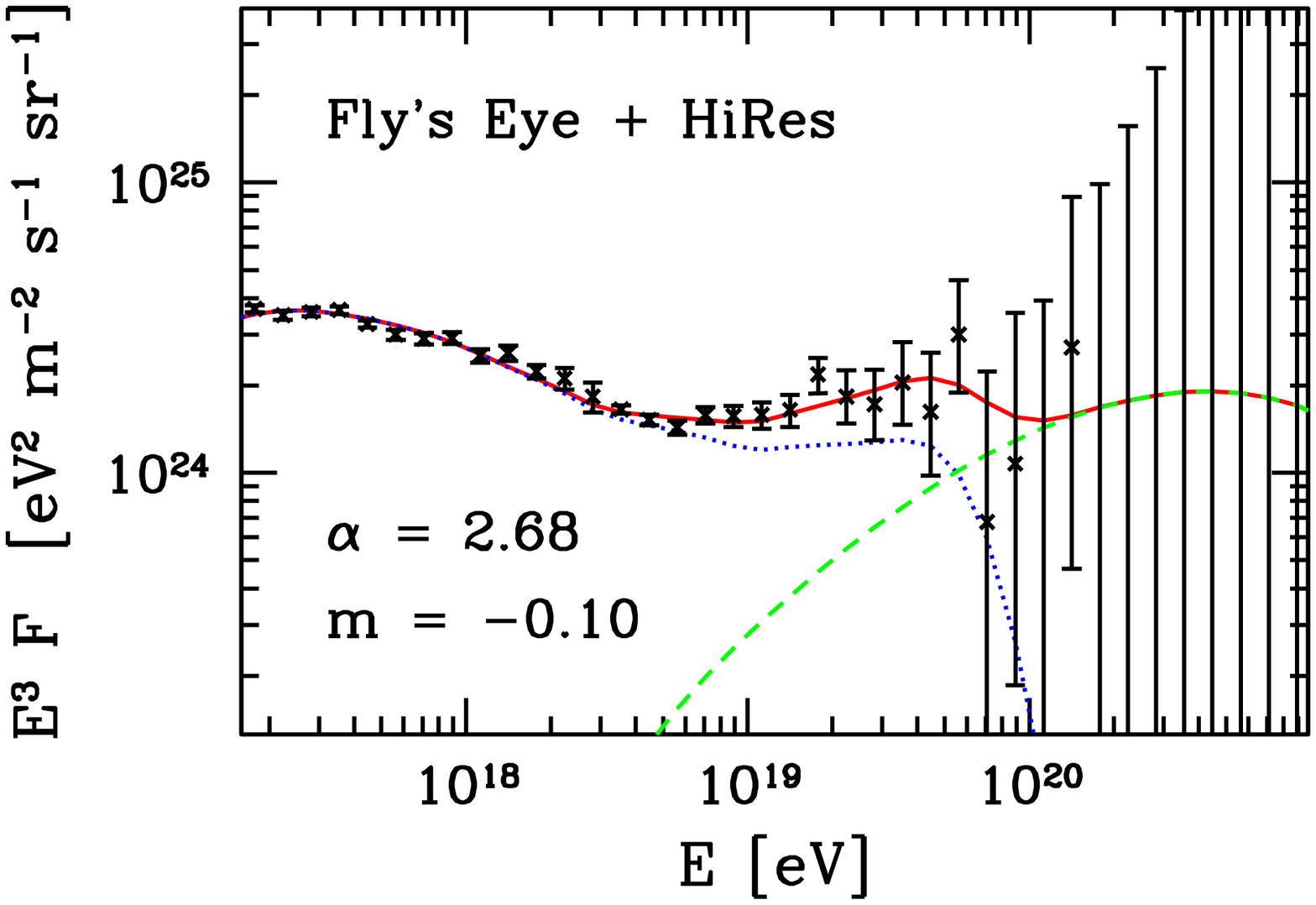}
\includegraphics*[width=8.3cm,clip=]{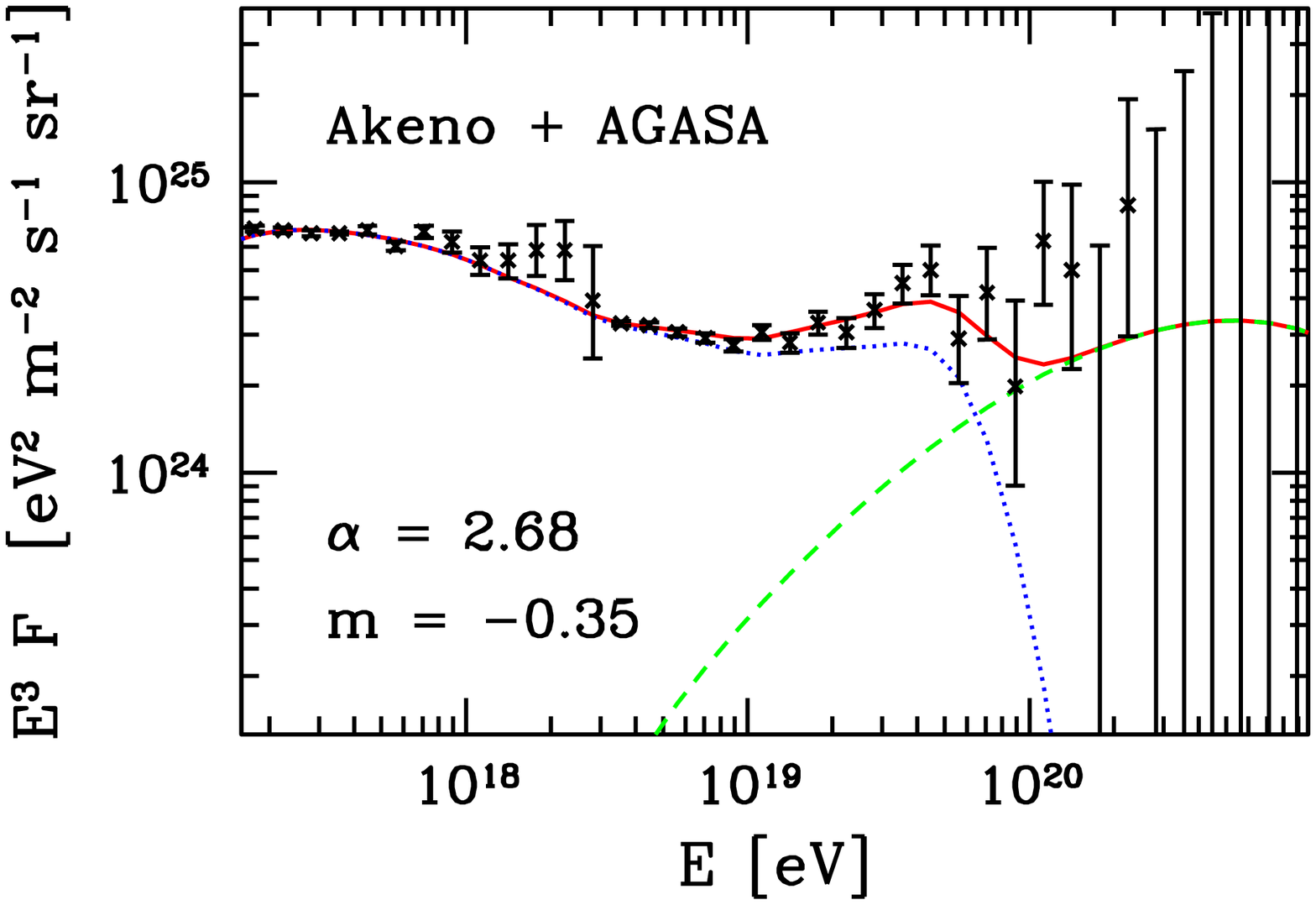}
\vspace{-1.5cm}
\caption[dum]{Ultrahigh energy cosmic ray data 
(points with statistical error bars; {\em left:} combination of Fly's Eye and HiRes data; 
{\em right:} combination of Akeno and AGASA data) 
and their best fits within the electroweak instanton scenario (solid) for 
$E_{\rm max}=3\cdot 10^{22}$~eV, $z_{\rm min}=0.012$, $z_{\rm max}=2$, consisting
of a proton component (dotted) plus a cosmogenic neutrino-initiated electroweak instanton-induced 
component (dashed).  
\label{fit}}
\end{center}
\end{figure}

{\it ad i)} In our comparison, we use the observed 
data from $\log(E/{\rm eV})=17.2$ to $\log(E/{\rm eV})=21$. 
We have altogether 38 bins. The 
bins with the largest energies are empty. This
non-trivial information is incorporated into the
analysis, too.  In the low energy region, there are no 
published results available from AGASA and only low statistics results
from HiRes-2.
Therefore, we included the results of the predecessor 
collaborations -- Akeno~\cite{Nagano:1991jz} and Fly's Eye, 
respectively -- into the analysis. With a small normalization correction,  
it was possible to continuously 
connect the AGASA data~\cite{Takeda:1998ps} with the Akeno ones and the 
HiRes-1 monocular data~\cite{Abu-Zayyad:2002ta
} with 
the Fly's Eye stereo ones~\cite{Bird:yi},
respectively (cf. Fig.~\ref{fit}). Usually, it is advisable to avoid the 
combination of different experimental data. Since in the 
present case it is 
interesting to see how well our scenario works for 
energies below and above the GZK cutoff, we used the less
problematic solution and combined results from experiments
with the same techniques and with largely overlapping 
experimental groups. The normalization was matched
at $\log(E/{\rm eV})=18.5$ for both cases.

Note that the highest energy event of HiRes was published using a five
times bigger bin size than for other energies~\cite{Abu-Zayyad:2002ta
}. 
In order to preserve information, we prefer  to keep the binning,  
give the particular bin with one event, and present 
upper bounds for the bins with zero event.  
From the published data of HiRes, we determined the 
approximate energy and the corresponding bin of the highest energy event.

{\it ad ii)} The goodness of the scenario is 
determined by a statistical analysis. 
In order
to give the confidence region in the $\alpha$--$m$
plane, we determined the compatibility of different
($\alpha ,m$) pairs with the experimental data.
For some fixed ($\alpha$,$m$) pair, one can
determine the expected number of event in 
individual bins (${\bf \lambda}=\{\lambda_1,...,\lambda_r\}$, where
the $\lambda_i$-s are non-negative, usually non-integer  
numbers and in our case the number of variables corresponds
to the number of bins, thus $r$=38). For this
specific ($\alpha ,m$), the probability
distribution in the $i$-th bin is given by 
the Poisson distribution with mean $\lambda_i$. 
The $r$ dimensional probability distribution
$P({\bf k})$ is just the product of the individual Poisson
distributions (here ${\bf k}=\{k_1,...k_r\}$ is a set 
of non-negative integer numbers).
It is easy to include also 
the $\approx 30\%$ overall uncertainty in the energy 
measurement of the experiments
into the $P({\bf k})$ probability. 
We denote the experimental result 
by ${\bf n}=\{n_1,...n_r\}$, where the $n_i$-s are
non-negative, integer numbers. According to the
$r$ dimensional probability distribution, the
experimentally observed event set 
${\bf n}$ has a definite, though
usually very small probability $P({\bf n})$. 
The ($\alpha , m$) pair is compatible with the 
experimental results if 
\begin{equation}\label{summation}
\sum_{{\bf k}|P({\bf k})>P({\bf n})}P({\bf k})<s\,.
\end{equation}
For a 1-(or 2-)sigma compatibility one takes s=0.68 
(or s=0.95), respectively. 
The best fit is found by minimizing the sum on the left hand side.
This technique is equivalent with the $\chi^2$
technique for a large class of 
problems\footnote{For the application
of the $\chi^2$ technique with UHECR data, see Refs.~\cite{Fodor:2001qy,
Fodor:2000za}.}. Note, however, that
the $\chi^2$ technique always gives a confidence region
and the $\chi^2/{\rm d.o.f}$ is used as an estimate for the goodness
of the scenario. Since $\chi^2/{\rm d.o.f}$ 
can be directly interpreted for
Gaussian problems  only, our goodness of the scenario
technique is more general. 

Since we have 38 variables, it is practically impossible to calculate
the sum in equation (\ref{summation}) exactly. Fortunately, there is no need for
the exact calculation, the sum can be determined
with arbitrary precision by using an importance sampling
based Monte-Carlo summation. Since the
sum of the individual probabilities is one, 
the left-hand-side of
equation (\ref{summation}) can be rewritten as 
\begin{equation}\label{monte-carlo}
\sum_{{\bf k}|P({\bf k})>P({\bf n})}P({\bf k})=
{\sum_{{\bf k}} P({\bf q})\, \theta[P({\bf k})-P({\bf p})]
\over \sum_{{\bf k}} P({\bf k})}\,.
\end{equation}
Equation (\ref{monte-carlo}) defines the Monte-Carlo 
summation straightforwardly. When calculating the sum, 
numbers with Poisson distribution are generated for
${\bf k}$ and only those are taken in the sum, for which
$P({\bf k})>P({\bf p})$. 

Figure~\ref{fit} shows our best fits for the HiRes and for 
the AGASA UHECR data (the lower energy data is also included, as
we explained before). The best fit values are $\alpha=2.68$, 
$m=-0.1$, for HiRes, and $\alpha=2.68$, $m=-0.35$, for AGASA.
We can see very nice agreement with the data
within an energy range of nearly four orders of magnitude. 
The fits are insensitive to the value of $E_{\rm max}$ as far as
we choose a value above $\approx 3\cdot 10^{21}$~eV.
The shape of the curve between $10^{17}$~eV and $10^{19}$~eV is mainly
determined by the redshift evolution index $m$. At $z=0$, below
$10^{18}$~eV
the attenuation length of protons is already around the size of the universe.
Therefore, one would expect no distortions of the injected spectrum below this energy
and an accumulation of particles just above it.
However, at larger redshifts, the interaction lengths are smaller and
the spectrum of particles created at cosmological distances has an accumulation
peak at lower energies. The more particles are created at large distances (i.e. 
the larger $m$ is), the stronger this effect will be\footnote{Our finding 
suggests that the extragalactic UHECR component begins to dominate over 
the galactic one already at 
$\approx 10^{17}$~eV. If we start our fit at $10^{18.5}$~eV -- assuming that 
the galactic component dominates up to this energy -- 
we find a very mild dependence on $m$ and the same best fit values for $\alpha$, 
with a bit larger uncertainties.}.
The peak around $4\cdot 10^{19}$~eV shows the accumulation of particles
due to the GZK effect. Neutrinos start to dominate over protons at around 
$10^{20}$~eV.

It is important to note that, if we omit the neutrino component, 
then the model is ruled out on the 3-sigma level for both experiments. 
This is due to the fact that there are no nearby sources 
($z_{\rm min} \neq 0$)
and all the events above $10^{20}$~eV are highly inconsistent with the 
predictions. A choice of $z_{\min}=0$ makes the HiRes data compatible with a 
proton-only scenario on the 2-sigma level 
(see also Refs.~\cite{Abu-Zayyad:2002ta,
Bahcall:2002wi
}). If neutrinos are included, then -- as they dominate over protons above $10^{20}$~eV -- 
the fit results cease to be sensitive to the value of $z_{\rm min}$.

\begin{figure}
\vspace{-1.6cm}
\begin{center}
\includegraphics*[width=8.3cm,clip=]{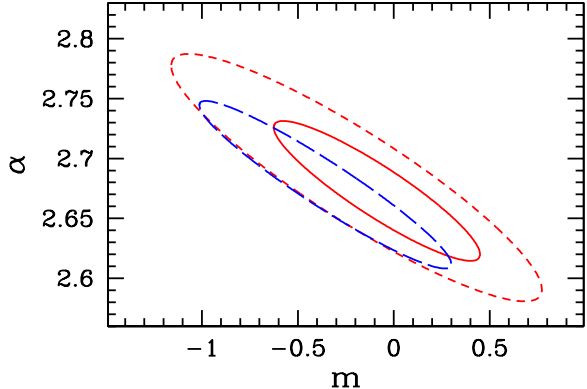}
\vspace{-1.5cm}
\caption[dum]{Confidence regions in the power-law index $\alpha$ 
and the redshift evolution index $m$ of the 
primary proton injection spectrum, for fits to the Fly's Eye + HiRes 
data (1-sigma (solid); 2-sigma (short-dashed)) and to 
Akeno + AGASA data (2-sigma (long dashed)), respectively, for 
$E_{\rm max}\gwig 3\cdot 10^{21}$~eV, $z_{\rm min}\geq 0$, $z_{\rm max}=2$.  
\label{regions}}
\end{center}
\end{figure}

Figure~\ref{regions} displays the confidence regions in the $\alpha -m$ plane
for HiRes and AGASA. The scenario is consistent on the 2-sigma level with
both experiments. For HiRes, the compatibility is even true on the 1-sigma level.
It is important to note that both experiments favor the same values for $\alpha$ and 
$m$, demonstrating their mutual compatibility on the 2-sigma level. 
If we ignore the energy uncertainty in the determination of the goodness 
of the fit, they turn out to be inconsistent.

Finally, let us discuss the consistency of our scenario with the currently
available limits on deeply penetrating showers from Fly's Eye~\cite{Baltrusaitis:mt}
and AGASA~\cite{Yoshida:2001icrc}. 
Taking into account -- in distinction to Ref.~\cite{Anchordoqui:2002vb} --  
the atmospheric attenuation  of the cosmogenic neutrino flux predicted in our scenario 
and the uncertainties in the estimate of the range of depth within which the shower must originate to
trigger the array, we find  that
AGASA should have seen $1\div 10$ quasi-horizontal air 
showers ($\theta \gwig 60^{\circ}$) from the electroweak instanton-induced processes
during a running time of $1710.5$~days. 
This is consistent with AGASA's present analysis of their respective data~\cite{Yoshida:2001icrc}.
The Fly's Eye upper limit on the product of the total neutrino flux times 
neutrino-nucleon cross-section, 
$(F_{\nu}\, \sigma_{\nu N}^{\rm tot})_{\rm Fly's\ Eye}$~\cite{Baltrusaitis:mt}, 
in the energy range $10^{17\div 20}$~eV, 
can be translated, for a given predicted neutrino flux $F_\nu^{\rm pred}$, 
into an upper limit on 
$\sigma_{\nu N}^{\rm tot}<(F_{\nu}\, \sigma_{\nu N}^{\rm tot})_{\rm Fly's\ Eye}/F_\nu^{\rm pred}$, 
as long as it is smaller than 
$10$~$\mu$b~\cite{Morris:1993wg,Tyler:2000gt
}. We find that, for our predicted cosmogenic neutrino flux, 
the right-hand-side of this inequality is larger than $10$~$\mu$b in the whole
energy range, such that the
Fly's Eye non-observation of quasi-horizontal air showers does not give any constraint. 
We therefore conclude that our prediction 
of the neutrino-nucleon cross-section, as shown in Fig.~\ref{cross-nuN} (left),
does not contradict any constraints from cosmic ray experiments 
so far, as long as the ultrahigh energy cosmic neutrino flux is at the
cosmogenic level we have predicted. 

\section{\label{conclusions}Summary and conclusions}

We have shown that a simple scenario with a single power-like 
injection spectrum of protons can describe all 
the features of the UHECR spectrum 
in the energy range $10^{17\div 21}$~eV. In this scenario, the injected protons
produce neutrinos during their propagation by interacting with the CMB.
Through Standard Model electroweak instanton-induced processes, these neutrinos
may interact with the atmosphere and give rise to a non-negligible contribution
to the detected air showers at the highest energies. The model has few 
parameters from which only two -- the power index $\alpha$ and the redshift  
evolution index $m$ -- has a strong effect on the final shape of the spectrum.
We found that for certain values of $\alpha$ and $m$ this scenario is 
compatible 
with the available observational data from the HiRes and AGASA experiments 
(combined with their predecessor  experiments, 
Fly's Eye and Akeno, respectively) on the 2-sigma level (also 1-sigma for
HiRes). 
The ultrahigh energy neutrino component can be experimentally tested by 
studying the zenith angle dependence of the events in the range 
$10^{18\div 20}$~eV at cosmic ray facilities such as the Pierre Auger 
Observatory and by looking for spatially compact energetic $\mu$ bundles
at neutrino telescopes such as AMANDA~\cite{Morris:1993wg}.  

Finally, let us emphasize that the same fit results are valid
for all strongly interacting neutrino models if the neutrino-nucleon 
cross-section has a similar threshold-like behaviour as in Fig.~\ref{cross-nuN}. 
The instanton scenario, however, has the advantage that it is based solely
on the Standard Model and can be falsified in the near future by 
a negative outcome of QCD instanton searches at HERA.

\section*{Acknowledgements}
We thank D.~V.~Semikoz for useful discussions about the
cosmogenic neutrino flux. This work was partially supported by Hungarian
Science Foundation grants No. OTKA-T034980/T037615.

\end{document}